\documentclass[prl,onecolumn,showpacs]{revtex4-2}
\usepackage{graphicx}
\usepackage{dcolumn}
\usepackage{amsmath}
\usepackage{amsfonts}
\usepackage{amssymb}
\usepackage[cp1251]{inputenc}
\usepackage[english]{babel}
\usepackage{graphicx}
\usepackage{bm}
\usepackage{color}
\usepackage{esvect}
\addto\captionsenglish{%
}
\usepackage{lettrine} 
\begin{document}



\title{Scanning quantum vortex microscopy reveals thickness-dependent pinning nano-network in superconducting Nb-films}


\author{Razmik~A.~Hovhannisyan$^{1}$}
\author{Sergey~Yu.~Grebenchuk$^{1}$}
\author{Andrey~G.~Shishkin$^{1,2,3}$}
\author{Artem~Grebenko$^{1}$}
\author{Nadezhda~E.~Kupchinskaya$^{1}$}
\author{Ekaterina~S.~Dobrovolskaya$^{1}$}
\author{Olga~V.~Skryabina$^{1,4}$}
\author{Alexey~Yu.~Aladyshkin$^{1,5}$}
\author{Igor~A.~Golovchanskiy$^{1}$}
\author{Alexey~V.~Samokhvalov$^{5}$}
\author{Alexander~S.~Melnikov$^{1,5}$}
\author{Dimitri Roditchev$^{6}$}
\author{Vasily S. Stolyarov$^{1,2,3,6}$}
\email{stolyarov.vs@phystech.edu}

\affiliation{$^1$Moscow Institute of Physics and Technology, 141700 Dolgoprudny, Russia}
\affiliation{$^2$ Dukhov Research Institute of Automatics (VNIIA), 127055 Moscow, Russia}
\affiliation{$^3$National University of Science and Technology MISIS, 119049 Moscow, Russia}
\affiliation{$^4$Institute of Solid State Physics RAS, 142432 Chernogolovka, Russia}
\affiliation{$^5$Institute for Physics of Microstructures RAS, 603950 Nizhny Novgorod, Russia}
\affiliation{$^6$LPEM, UMR-8213, ESPCI Paris, PSL, CNRS, Sorbonne University, 75005 Paris, France}

\date{\today}

\maketitle

\section{Abstract}


\lettrine{T}{}he presence of quantum vortices determines the electromagnetic response of superconducting materials and 
devices. Controlling the vortex motion, their pinning on intrinsic and artificial defects is therefore essential for superconducting electronics. Here we take advantage of the attractive force between a magnetic cantilever of the Magnetic Force Microscope and a single quantum vortex to spatially map the pinning force inside 50-240 nm thick magnetron-sputtered Nb-films, commonly used in advanced superconducting electronics. The revealed pinning nano-network is related to the thickness-dependent granular structure of the films as well as to the characteristic microscopic scales of superconductivity.  Our approach is general, and can be directly applied to other type II granular superconducting materials and nanodevices.

\section{Introduction}

Defects play a crucial role in
superconductivity~\cite{menard2015coherent,stolyarov2018domain,grebenchuk2020crossover,stolyarov2022effective,vagov2023intertype}. In superconducting electronic devices, their presence is often detrimental or
unwanted~\cite{Larbalestier_2018,golod2021reconfigurable,grebenchuk2020observation,grebenchuk2022magnetic,golod2022demonstration}. In other cases, such as superconducting wires and cables, they enable pinning Abrikosov vortices thus enhancing the critical current density \cite{Golovchanskiy_JAP_114_163910, Golovchanskiy_ASS_311_549}. Moreover, disordered superconducting films have high kinetic inductance~\cite{Annunziata_2010,carbillet2020spectroscopic}, making them promising for applications in superconducting quantum devices and sensors~\cite{Soloviev_2021}.

The physics of the vortex-defect interaction in type-II superconductors is also a very important fundamental problem~\cite{Campbell1972,Blatter1994}.
For several decades different scenarios of vortex pinning have been studied including pinning on columnar defects~\cite{BezryadinPLA1994,BezryadinJLTP1995,BerdiyorovEPL2006,BerdiyorovPRL2006,BerdiyorovPRB2006,SabatinoJAP2010,LatimerPRB2012,LatimerPRL2013,GePRB2017,XueNJP2018}, blind holes~\cite{BezryadinPRB1996,BerdiyorovNJP2009}, non-superconducting inclusions~\cite{KarapetrovPRL2005,SadovskyyPRB2017,WillaSUST2018}, among others (see Ref.~\cite{KwokRPP2016} for a review). Basic mechanisms of the vortex trapping on defects are related either to the change in the energy of supercurrents~\cite{BeanPRL1971} or to the changes of the vortex core energy~\cite{Campbell1972,Blatter1994}. The progress in technology enables tuning and controlling the vortex pinning using the sample thickness modulation~\cite{DaldiniPRL1974}, substrate engineering~\cite{CrisanAPL2001}, surface decoration with magnetic nanoparticles~\cite{MartinPRL1997,VanBaelPRB1999}, engineering of the specific pinning centers ~\cite{FeighanSUST2017}, and ion irradiation~\cite{BugoslavskyN2001,NakajimaPRB2009,ZechnerSUST2018,AntonovPSS2019,AntonovPSS2020,AntonovPC2020,AichnerFNT2020}.

A detailed knowledge of the pinning network parameters is deeply desired in all cases.
However, their experimental studies are very challenging, as even tiny non-magnetic defects such as grain boundaries or non-superconducting inclusions could serve as an efficient pinning
centres on the scale of the superconducting coherence length. This covers spatial scales ranging from a few nanometers to several
microns. Thus, an ideal probe should have a nano-scale resolution combined with a large field of view; it should probe bulk
properties while being non-destructive.

Several microscopies enable nanometer-scale defect imaging. Transmission electron microscopy offers the analysis down to atomic scale \cite{Zheng_2015, Devred_2004,kudriashov2022revealing,golovchanskiy2023magnetization} but is destructive and probes a tiny part of the sample. Scanning methods such as electron~\cite{martinez2013microstructures}, tunneling~\cite{Lee_2011,stolyarov2018expansion,berti2023scanning,koslowski2004studying} and atomic force microscopy~\cite{wu2005studies,Volodin_2002} also show an excellent spatial resolution and are non-destructive. Yet, they only reveal defects that protrude at the surface (e.g. cracks or grain boundaries) and provide only limited information about the defect distribution in the bulk. More dedicated methods probe specific superconducting properties. Magneto-optical imaging
\cite{Vlasko_1992,Koblischka_1995,Runge_2000,Veshchunov_2016},
Lorentz microscopy \cite{Tonomura_1992}, magnetic decoration
\cite{Vinnikov_2012,Vinnikov_2019}, scanning SQUID
\cite{Kirtley_2016,Embon_2017,finkler2012nano,halbertal2016nanoscale,Embon_2015}, scanning Hall probe
\cite{Tamegai_2001,Kalisky_2009,Park_2012,Moschalkov_2015} and
magnetic force microscopy (MFM)
\cite{Volodin_2002,Auslander_2009,Budakian_2019,Dremov_2019,Correa_2019} probe spatial variations of the magnetic field outside the
sample and reflect the distribution of screening (Meissner) and
transport currents in the material. Low-temperature scanning laser
\cite{Zhuravel_2006,Koelle_2017,Galin_2020} and scanning electron
microscopy \cite{Clem_1980,Doenitz_2007,Rosticher_2010} probe
thermal healing processes within a superconductor subject to local heating by the beam~\cite{Clem_1980}. These techniques are non-destructive, access the bulk properties and thus provide useful information about the properties of superconducting cables~\cite{Vlasko_1992,Koblischka_1995,Runge_2000,Volodin_2002,Auslander_2009,Park_2012,Embon_2017} and devices~\cite{Zhuravel_2006,Doenitz_2007,Kalisky_2009,Rosticher_2010,Giovati_2012,Koelle_2017,Dremov_2019,Galin_2020}.
However, these methods have spatial resolution typically in the range of microns, thus missing nanoscale defects. Therefore, the quest for a high-resolution, non-destructive method of defect network characterization in superconductors is still open. 

\begin{figure*}[t!]
\begin{center}
\includegraphics[width=18 cm]{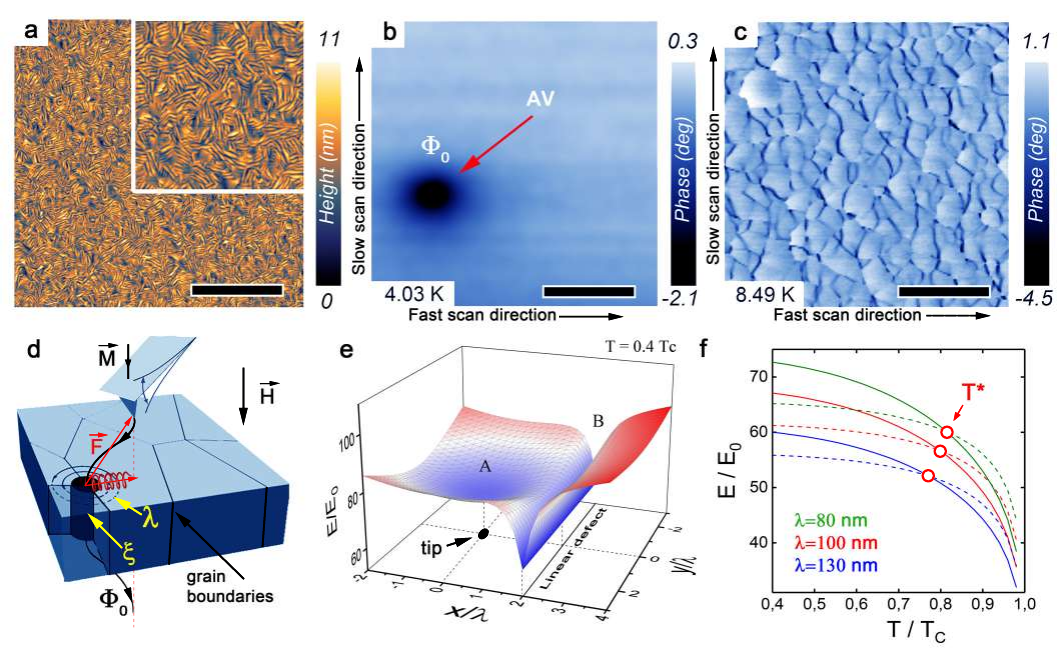}
\caption{\textbf{Principles of SQVM.}  \textbf{a} -- Room-temperature AFM $3\times3$ $\mu m^2$ image of a 100 nm thick Nb film. The inset represents the $0.5\times0.5$ $\mu m^2$ zoom on the sample surface. \textbf{b,c} -- MFM (b) and SQVM (c) images of 100 nm thick film acquired at $T=4.03$ and 8.49\,K, respectively. The black scale bar in (a-c) corresponds to 1 $\mu$m. \textbf{d} -- sketch of SQVM experiment. At low temperatures, $T \ll T_c$, the vortex pinning is stronger than the vortex attraction by the cantilever. \textbf{e} --  calculated variation of the normalized free energy as a function of the vortex position at $T= 0.4\, T_c$ (see definitions in the text). The cantilever apex is located in A and the linear defect in B. \textbf{f} -- temperature dependence of the normalized free energy for the vortex located under the cantilever in $A$ (solid lines) and at the defect in $B$ (dashed lines). Blue, red, and green curves correspond to the samples of the thickness $d$ = 50 nm, 100 nm and 240 nm, respectively (see Section Methods for further information). The crossover temperature $T^*$ (red open circles) depends on film thickness $d$.}
\label{Fig.1}
\end{center}
\end{figure*}

In the present work we study the spatial and temperature evolution of the vortex pinning in magnetron-sputtered 50-240 nm thick Nb-films which are widely used in superconducting electronics and quantum technology~\cite{fietz1969hysteresis,park1992vortex,dasgupta1978flux}. We use the magnetic cantilever of the Magnetic Force Microscope to generate a single quantum vortex in the studied sample upon its cooling below the superconducting critical temperature. Once created, the vortex is attracted to the apex of the cantilever. During the cantilever scanning over the studied region of the film, the dragged vortex explores the superconductor by jumping from one pinning center to the other. These successive jumps are detected through the modifications of vortex-cantilever force \cite{Auslander_2009} and are presented in form of spatial maps. Since the vortex pinches through the whole thickness of a superconductor, it is the bulk pinning potential that is probed, and its spatial distribution projected to 2D-maps is obtained. In the studied Nb-films, the revealed pinning network is related to the granular structure of the films. Surprisingly, the spatial resolution $\sim$20 nm of this non-destructive Scanning Quantum Vortex Microscopy (SQVM) is found to be much higher than the expected limitation $\sim$250 nm due to the lateral extent of the vortex-cantilever magnetic interaction. The analysis of the data shows that the SQVM resolution is related to the superconducting coherence length, and that the microscopic origin of the pinning is the vortex core blocking by the grain boundaries.

\
\section*{Results}
\
The studied Nb films were deposited onto the silicon substrate using a standard magnetron sputtering (see Section Methods for further details). The properties of these commonly used films have been extensively studied in the past. Several works focused on the vortex pinning on both intrinsic and artificial defects~\cite{park1992vortex,golod2021reconfigurable}. The films are known to exhibit a strong vortex pinning on structural defects formed during deposition~\cite{pinto2018dimensional}. 

Fig.~\ref{Fig.1}(a) represents  $3\times 3$ $\mu m^2$ atomic force microscopy image realized at room temperature on the surface of 100 nm thick film. The inset is a  $0.5\times 0.5$ $\mu m^2$ zoom on the same sample area. The granular structure of the film is clearly revealed; the grains appear elongated with the apparent length $\sim$ 30-50 nm and width $\sim$ 5 nm. The grain boundaries are tiny and are not resolved. Remarkably, the neighboring grains are co-aligned forming larger agglomerates $\sim$ (30-50 nm)$^2$ separated by larger voids appearing as dark spots. 

To realize the vortex pinning maps by SQVM, the first step is to create an interacting probe - a single Abrikosov vortex, in the present case. The samples were put in the cryogenic MFM (see Section Methods) and cooled below the critical temperature of the superconducting transition $T_c \simeq$ 9.05 K. During the cooling process, the Co/Cr magnetic cantilever of the MFM was kept above the sample surface at a distance (lift) $\sim 2$ $\mu m$. At this lift, the stray magnetic field of the cantilever threading the sample is only a few Oe; though, this field is enough to create one or a few magnetic flux quanta in the area of interest. Below $T_c$, this magnetic flux becomes quantized in the form of a quantum vortex. Note that in general, even zero-field cooled samples may freeze some quantum vortices due to the Earth field. While these vortices can also be used for SQVM, their initial location is uncontrolled.


Fig.~\ref{Fig.1}(b) displays a $3\times 3$ $\mu m^2$ MFM map of the film acquired in the area where the quantum vortex was expected to be created at $T=$4.03 K (that is far below $T_c^{100nm}$= 9.0 K) and cantilever lift 200 nm. Note that in all presented images, the horizontal axis follows the direction of the fast scan. In this image, the grey contrast represents the phase shift (grey-scale coded) of the MFM cantilever oscillations. Since the cantilever oscillates in the direction perpendicular to the film surface, the phase shift is proportional to the normal component of the force experienced by the cantilever: it is positive for the repulsive normal forces and negative for the attractive ones. Due to the Meissner diamagnetism, the main interaction between the cantilever and the film in the absence of a vortex is repulsive; it is witnessed by a positive phase shift measured on most of the scanned area. However, a spot - a single Abrikosov vortex - is observed near the image centre. This vortex was created by the cantilever during cooling when the sample became superconducting. The spot is black (negative phase shift) since the interaction between the cantilever and the vortex created by itself is attractive. The reason why the vortex appears in this image and does not follow the moving cantilever despite the vortex-cantilever attraction is that at the present experimental conditions (temperature, lift, tip magnetisation), the vortex pinning force by the sample exceeds its attraction by the cantilever, thus fixing the vortex position.

The MFM map presented in  Fig.\ref{Fig.1}(c) was realized at $T=$8.49 K, that is 0.5 K below $T_c^{100nm}$; the same sample region as in Fig.\ref{Fig.1}(b) was explored. The overall gray background on this map corresponds to the positive phase shift, yet the diamagnetic repulsion is slightly larger than in Fig.~\ref{Fig.1}(b), due to a lower cantilever lift $h = 80$ nm used. On this map, no pinned vortex is visible anymore. Instead, a remarkable fish-skin-like pattern is observed with a nano-network of sharp dark boundaries where a strong attractive force is registered. Since the vortex is topologically robust phase singularity, and cannot disappear, a tentative interpretation of this observation is that at these experimental conditions the vortex gets unpinned and dragged by scanning cantilever, thus exploring the pinning potential of the sample. In locations corresponding to phase drops, the moving cantilever exercises a stronger force to unpin and drag the vortex and therefore, in these locations the pinning is stronger. Thus, the phase shift maps of this Single Quantum Vortex Microscopy (SQVM) reveal the spatial distribution of the pinning potential and its local strength. This is the central observation of the present work.

The geometry of SQVM is sketched in Fig.~\ref{Fig.1}(d); the theoretical justification of the approach is provided in Figs.~\ref{Fig.1}(e,f) (see Section Methods). At the superconducting transition, a vortex is created using the stray field $\vec{M}$ of the magnetic cantilever. The vortex has a core of the size of the thickness-dependent coherence length $\xi_d$ ($\xi_d \sim$ 10-20 nm in the studied films); the vortex supercurrents circulate around the core on the scale of the thickness-dependent magnetic penetration depth $\lambda_d > \xi_d$. The interaction force $\vec{F}$ of such a vortex with the cantilever has a magnetic origin and can be seen as effectively attractive. Indeed, in the absence of the vortex, the magnetic flux from the oscillating cantilever is screened by Meissner currents resulting in a repulsion, while in the presence of a vortex, the vortex currents circulating in the direction opposite to the Meissner ones reduce the repulsion. The plot in Fig.~\ref{Fig.1}(e), obtained at the conditions close to the experimental ones in Fig.~\ref{Fig.1}(b), depicts the free energy of the system as a function of the vortex location, for fixed positions of the cantilever, in A, and of the linear defect, in B. The energy is normalized to $E_0 = \Phi_0^2 / 64 \pi^3\lambda_d(0)$, where $\Phi_0=h/2e$ is the flux quantum, $\lambda_d(0)$ is the zero-temperature magnetic penetration depth taken equal to 130 nm, 100 nm and 80 nm  for, respectively, $d$ = 50 nm, 100 nm, and 240 nm thick films~\cite{Gubin}. The free energy has a local minimum when the vortex is located below the cantilever. When the cantilever in A and vortex get misaligned, the energy increases progressively on the lateral scale of the magnetic penetration depth, resulting in an increasing pooling force presented by a red spring in Fig.~\ref{Fig.1}(d). The other minimum exists in B, when the vortex sits at the pinning defect. In the simulation  Fig.~\ref{Fig.1}(e), the interaction that pins a vortex at the linear defect is stronger than the shallow minimum at the cantilever position in A. At this condition, the vortex remains strongly pinned at the defect, enabling its visualisation in the MFM experiment, Fig.~\ref{Fig.1}(b). However, the temperature evolution of the two minima (in A and in B) are different, as demonstrated in Fig.~\ref{Fig.1}(f). At low temperatures, the energy of the system with a vortex pinned at the linear defect in B can indeed be significantly lower that the energy with the vortex in A. Close enough to $T_c$ the situation inverses. In terms of forces it means that the maximum pooling force of the cantilever now exceeds the pinning one. At this new condition, the scanning cantilever will unpin the vortex from the defect and drag it. This situation is realized in the SQVM experiment, Fig.~\ref{Fig.1}(c). The crossover temperatures $T^*$ between the two regimes are presented by red open circles in Fig.~\ref{Fig.1}(f). The existence of $T^*$ is confirmed experimentally (see Supplementary Information SFig.1). Note, that the simulations in Fig.~\ref{Fig.1}(f) predict the crossover temperature to depend on film thickness.

The above considerations  suggest that to enable the SQVM experiment the vortex has to be unpinned from the defect and follow the scanning cantilever. This requires the vortex-cantilever effective attractive interaction to exceed the pinning potential. The attraction can be increased by reducing the lift, while the pinning can be lowered by raising the temperature towards $T_c$. Indeed, the considerations behind the energy plots, Fig.~\ref{Fig.1}(e,f), take into account the vortex currents that circulate around the core and explore the disorder potential on the scale of the penetration depth $\lambda_d$ as well as on the energy of the magnetic flux the currents create. Moreover, on the microscopic level, at least a part of the pinning potential is related to the energy $E_{core}$ of the vortex core whose lateral size is of the order of $\xi_d$. This energy is positive, due to the suppression of the superconducting order parameter (gap) $\Delta(r)$ inside the core: $E_{core} \sim N(E_F) \Delta^2 \times \xi_d^2 d$, where $\frac{1}{2} N(E_F) \Delta^2$ is the condensation energy density, $N(E_F)$ is the density of electronic states at the Fermi level $E_F$, and $\xi_d^2 d$ is the volume occupied by the core in the film. At a non-superconducting defect $\Delta(r) \rightarrow 0$ and, if the defect has a size $l \sim \xi_d$ and a substantial height $\sim d$, the energy $E_{core}$ is reduced if the core coincides with the defect. That is why such defects (and particularly columnar ones) are usually strong vortex pinning centres. At $T \rightarrow T_c$ both $\xi_d, \lambda_d \rightarrow \infty$. Thus, close enough to $T_c$, the vortex core and the vortex currents occupy the areas much larger than the size of individual defects; this "averaging over disorder" leads to the smearing of the pinning potential and to the consequent reduction of the pinning force, thus enabling SQVM experiment in the temperature range $T^*<T<T_c$. In addition, the thermal fluctuations and reduced Josephson inter-grain coupling also contribute to the depinning at higher temperatures.

\begin{figure*}[ht!]
\begin{center}
\includegraphics[width=18cm]{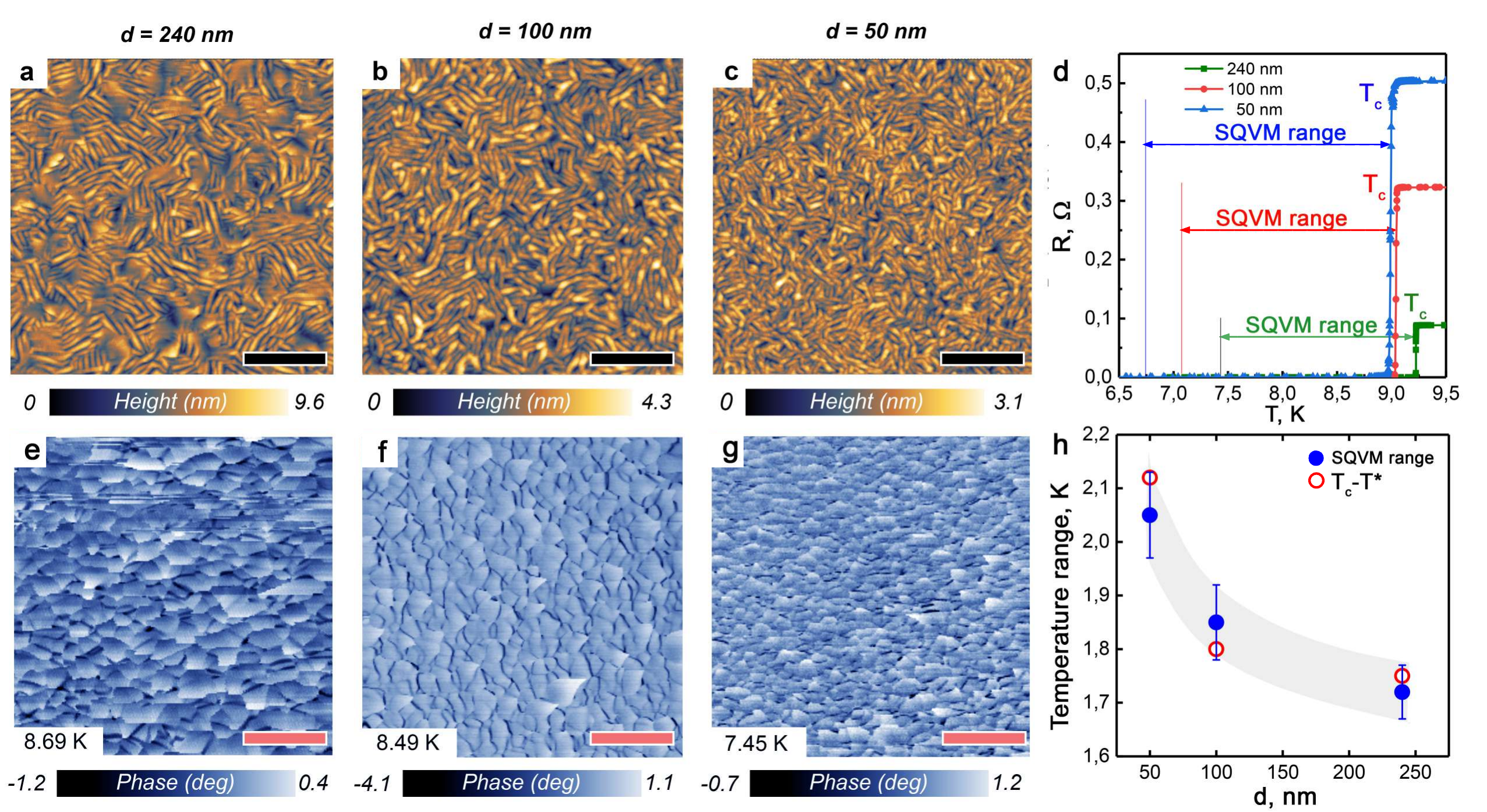} 
\caption{\textbf{Thickness dependent vortex pinning network in Nb-films.} \textbf{a}-\textbf{c} -- 1$\times$1 $\mu$m$^2$ room-temperature AFM images of 240 nm, 100 nm and 50 nm thick Nb films, respectively. The black bars correspond to 200 nm. Both grain size and apparent surface roughness increase with thickness. \textbf{d} -- Resistance of the films measured near the superconducting transition by the four-probe transport. For each thickness, colored arrows visualize the temperature range where SQVM experiment was enabled. \textbf{e}-\textbf{g} - 5$\times$5 $\mu$m$^2$ SQVM maps (acquired at $h$= 80 nm) of the films showed in \textbf{a}-\textbf{c}. The orange bars correspond to 1 $\mu$m. \textbf{h} - Blue circles: dependence of the SQVM temperature range on film thickness. Red open circles: calculated $T_c-T^*(d)$ (see section Methods).}
\label{Fig.2}
\end{center}
\end{figure*}

To ensure that the network presented in Fig.~\ref{Fig.1}(c) is indeed related to the Nb-film structure, further SQVM experiments were provided on Nb-films of different thicknesses. In Figs.~\ref{Fig.2}(a-c), room-temperature topographic AFM images of the films are presented. They demonstrate the expected granular structure and the well-known increase of the Nb-grain size with increasing film thickness. The superconducting properties of the films were characterized by four-probe electron transport; the results are presented in Fig.~\ref{Fig.2}(d). The high film quality is witnessed by a high $T_c^{240nm}\approx$~9.2~K for 240 nm thick film; the expected slight decrease of $T_c$ for thinner samples is also observed. The corresponding SQVM maps in Figs.~\ref{Fig.2}(e-g) also show the same tendency: the characteristic spatial scales of the observed pinning network grow with the film thickness and thus clearly correlate with the grain size. Notably, the observed spatial networks barely depend on temperature (see Supplementary Figure S2), witnessing for its direct relation to the film structure.

\section*{Discussion}
\

We now discuss a deep connection of the SQVM to the film properties. We start with the analysis of the temperature range close to $T_c$ where the SQVM experiments are enabled (showed by colored arrows in Fig.~\ref{Fig.2}(d) and plotted as blue circles in Fig.~\ref{Fig.2}(h)). Experimentally, the SQVM temperature range was found to significantly vary upon the film thickness: It gets wider with decreasing $d$ (blue circles in Fig.~\ref{Fig.2}(h)). Remarkably, this dependence almost perfectly coincides with the calculated $T_c-T^*(d)$, taking $T^*(d)$ from the numerical simulation presented in Fig.~\ref{Fig.1}(f). Note that the dependence is not trivial: to achieve a good agreement presented in Fig.~\ref{Fig.2}(h), calculations required accurately considering experimentally established thickness-dependent $\lambda_d$~\cite{Gubin}. Therefore, by measuring the SQVM range and by inverting the used numerical protocol, it is possible to extract the effective $\xi_d$ and $\lambda_d$. This is of a high interest for ultra-thin superconducting films in which these characteristics are not straightforward to measure directly.

\begin{figure*}
\begin{center}
\includegraphics[width=18cm]{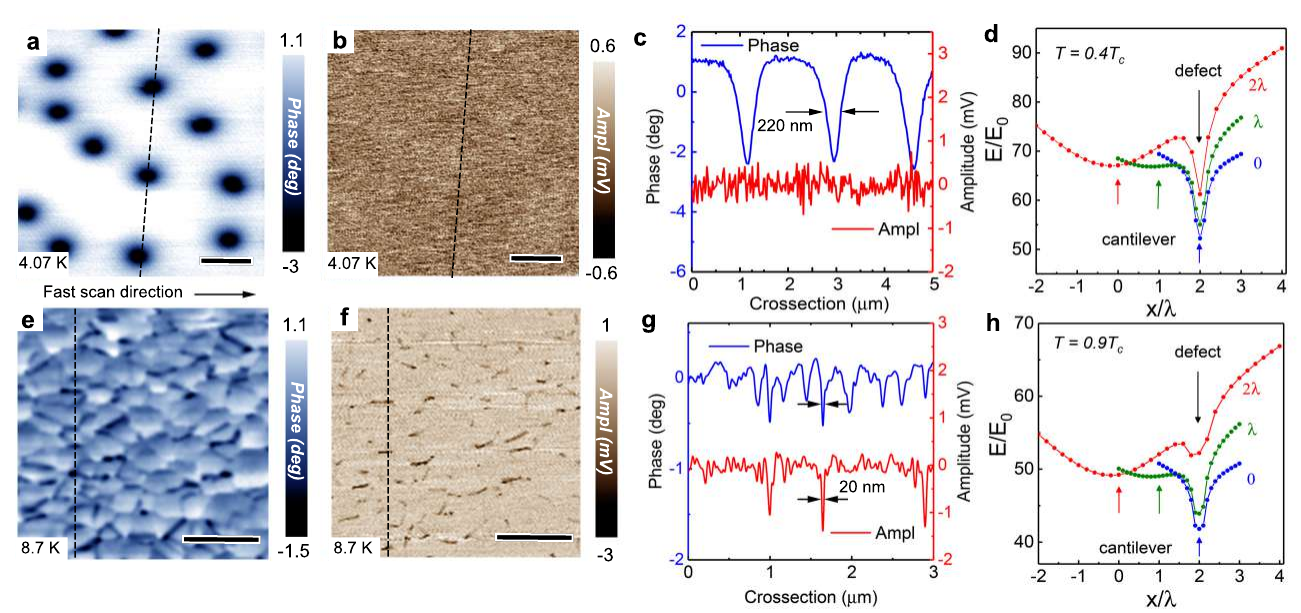} 
\caption{ \textbf{From pinned vortex MFM imaging to SQVM.} \textbf{a, b} -- $3\times 3 \mu \text{m}^2$ MFM phase (a) and oscillation amplitude (b) maps acquired at $T=$ 4.07 K and $h=$ 80 nm in the 10 Oe field-cooled 240 nm thick Nb-film. \textbf{c} -- cross-section plots of the phase (blue curve) and oscillation amplitude (red curve) following the dashed lines in maps \textbf{a, b}. \textbf{d} -- calculated free energy at 0.4 $T_c$ as a function of cantilever position (see section Methods). \textbf{e, f} -- SQVM phase and amplitude maps of the same sample acquired at $T=$ 8.7 K and $h=$ 80 nm. \textbf{g} -- a cross-section of the phase (blue curve) and amplitude (red curve) maps in \textbf{e, f}. The black bars in \textbf{a,b,e,f} correspond to 1 $\mu$m. \textbf{h} -- calculated free energy at 0.9 $T_c$ as a function of cantilever position (see section Methods).
}
\label{Fig.3}
\end{center}
\end{figure*}

We now focus in more detail on the interaction of vortices with the cantilever - a key for understanding SQVM. In the experiment presented in Fig.~\ref{Fig.3}(a-c), we cooled a 240 nm thick film down to 4.07 K (that is well below its $T_c=$ 9.2 K) in the presence of an external magnetic field $\vec{H}$=10~Oe aligned with the stray field $\vec{M}$ of the cantilever (Fig.~\ref{Fig.1}(d)). On the phase map of Fig.~\ref{Fig.3}(a) obtained at $T=4.07$ K (that is well below the SQVM temperature range, Fig.~\ref{Fig.2}(d,h) ), several vortices are visible forming a disordered vortex lattice, as expected; all vortices have the same apparent size. The section of the phase map along the dashed line in Fig.~\ref{Fig.3}(a) is presented as a blue curve in Fig.~\ref{Fig.3}(c). It demonstrates the "vortex size", about 220 nm at half maximum (at 80~nm lift). Such a large value is due to the magnetic interaction between the vortex field laterally extending to $\sim \lambda_d$, and the cantilever field extending over hundreds of nanometers~\cite{Hovhannisyan_2021}. At the same time, the cantilever oscillation amplitude map presented in Fig.~\ref{Fig.3}(b) is featureless (see also the section plot in Fig.~\ref{Fig.3}(c, red line). In fact, the cantilever oscillation amplitude is related to the dissipation in the coupled system vortex-cantilever~\cite{Dremov_2019,stolyarov2022revealing}. At the temperature of the experiment $\approx$ 0.4~$T_c$, the theoretical curves in Fig.~\ref{Fig.3}(d) obtained for different cantilever positions with respect to the vortex location (above the defect (blue line), at a distance of $\lambda_d$ (green line) and 2$\lambda_d$ (red line)) demonstrate that the cantilever is unable to unpin and drag vortices. The vortices remain pinned and do not dissipate. That explains why the dissipation-related amplitude map in Fig.~\ref{Fig.3}(c) is featureless.

The SQVM map taken at $T=$ 8,7 K ($T \simeq 0.9~T_c$) is presented in Fig.~\ref{Fig.3}(e). On this phase map, individual vortices are not visible anymore but the pinning network is (similarly to the result presented in Fig.~\ref{Fig.1}(c) ). In this case, sharp drops are also visible on the simultaneously recorded amplitude map, Fig.~\ref{Fig.3}(f); the position of the drops spatially coincide (the spatial correlation between amplitude and phase signals is clear in Fig.~\ref{Fig.3}(g) ). The calculations presented in Fig.~\ref{Fig.3}(h) show that at this high temperature, the attraction by cantilever exceeds the pinning force: the vortex gets unpinned and dragged, thus enabling SQVM experiment. The dissipation in the system vortex-cantilever varies upon the location of the vortex that explores the pinning network.


One of the puzzling features in the SQVM maps is therefore their surprising spatial sharpness: the phase drops occur on the scale $\sim$ 15-30 nm (see plots in Fig.~\ref{Fig.3}(g)). This is by far shorter than the characteristic scales of sample-cantilever magnetic interaction ($\lambda_d (T) \gtrsim$ 100 nm, tip size, tip lift, etc.), which presumably would limit the spatial resolution of SQVM. To resolve the puzzle, one should recall the basic principles of MFM. The magnetic cantilever represent a mechanical mass-spring oscillator with the resonant frequency $\omega_0$. In the experiment, this oscillator is excited by a piezoelectric dizzier at a close frequency $\omega$~\cite{Dremov_2019}. The tip oscillates in the direction perpendicular to the surface; its position is $z=z_0 \cos (\omega t+\theta)$. In the presence of an external force with a non-zero $z$-component of the force gradient, the frequency $\omega$, the oscillation amplitude $z$ and the phase shift $\theta$ change by \cite{Mironov_2004,di2019quantitative}:

\begin{equation}
\delta \omega \approx \frac{\omega}{2k}\frac{\partial F}{\partial z}, \; 
\;
\delta z \approx \frac{2z_{0}}{3\sqrt{3}}\frac{Q}{k}\frac{\partial F}{\partial z}, \; \;
\delta\theta\approx\frac{Q}{k}\frac{\partial F}{\partial z}, \; \;
\end{equation}

where $Q=\frac{k z_0^{2}\omega_0}{2 P_\mathrm{dis}}$ is the quality factor of the cantilever, and $P_\mathrm{dis}$ is the dissipated power. The vortex is strongly pinned at low temperatures, and the dissipation is low (no vortex core motion, no quasiparticles). At these conditions, the vortex appears in $\delta\theta(x,y)$ maps (as in Fig.~\ref{Fig.1}(b), for instance) due to the spatial variations of the force gradient $\frac{\partial F}{\partial z}$. When the cantilever apex is located away from a vortex, the interaction force $F$ is mainly due to the repulsion by Meissner currents of the sample attempting to screen the stray magnetic field of the cantilever. Though, an attractive vortex-cantilever force dominates when the cantilever is situated close or above the vortex. Both vortex-generated and Meissner current-generated forces decay with increasing the cantilever-surface distance, but their directions are opposite, and thus, also are their gradients. That is why the phase shifts in Fig.~\ref{Fig.1}(b) measured above the vortex and away from it have opposite signs. Note also that the attractive vortex-cantilever force is on the origin of the vortex drag when the tip is moved away from the vortex centre. At low temperatures, this force does not exceed the pinning force; the vortex remains pinned enabling its imaging. The apparent vortex size of a few hundreds of nanometers (Figs.~\ref{Fig.1}(c),~\ref{Fig.3}(a,c) results from the convolution of the lateral extend of the vortex field $\sim \lambda_d ($4K$)\simeq$ 100 nm \cite{Gubin} with the magnetic footprint of the cantilever~\cite{Hovhannisyan_2021}.
When the temperature is increased, $\xi_d(T)$ and $\lambda_d (T)$ raise, the pinning force decreases and, at some temperature, becomes lower than the vortex-cantilever attraction. The vortex gets unpinned and can be dragged by the cantilever, as depicted in Fig.~\ref{Fig.1}(d). Close enough to $T_c$, vortex-cantilever attraction strongly dominates, and the vortex can be seen as rigidly "attached" to the scanning cantilever, while the vortex core and vortex currents interact with the local pinning network. At $T$= 8.5~K of the SQVM experiment presented in Fig.~\ref{Fig.1}(c), the vortex currents are already spread over large distances $\gtrsim 2\lambda ($8.5~K$) \sim$ 500~nm; they interact with a large number of local pinning centres (point defects, grain boundaries, etc.). Therefore, at this temperature one does not expect sharp spatial variations in $\delta\theta(x,y)$ maps due to spatial variations of the vortex current distribution. However, the vortex core has a much smaller size $\sim  \xi$ estimated to 15~nm at $T$=4~K~\cite{Zeinali_2016}, and to 38~nm at $T$=8.5~K. The core motion is dissipative; the dissipation depends on the normal state resistance of the material in the place occupied by the core. The grain boundaries are composed of disordered and partially oxidized Nb; the vortex core motion along and across these grain boundaries is more dissipative as compared to that in superconducting Nb. At such defects $P_\mathrm{dis}$ raises, the quality factor $Q$ drops, resulting in sharp phase/amplitude shifts in SQVM maps. Local variations of $P_\mathrm{dis}$ are tiny but owing a very high $Q$-factor of the cantilever, $Q\sim$4000, the detection of the vortex core motion is rather easy \cite{Dremov_2019,Hovhannisyan_2021,stolyarov2022revealing,grebenchuk2022magnetic}. Note that owing to the spatial sharpness of SQVM maps (Fig.~\ref{Fig.3}(g)), the position of the pinning centers can be determined with a high accuracy $\sim 1$~nm. This demonstrates a true nanoscale resolution of the SQVM we developed and used in this work.

In the present work we studied standard magnetron-sputtered Nb-films which are commonly used in superconducting technology such as rapid single flux quantum electronics, qubits, registers, sensors, etc. In these applications, the precise knowledge of local superconducting properties is strongly desired. With this respect, the implemented SQVM is a new powerful tool as it provides a direct information on vortex pinning with a nano-scale resolution limited by the superconducting coherence length; in commonlgy used type-II superconductors (Nb, NbN, TiN,..) the latter is much shorter than the magnetic penetration depth. In the present case of thin Nb-films, the SQVM shows that there is no direct correlation between granular structure of the films and their local superconducting properties and thus, the knowledge of the film morphology is not enough to decide on superconducting properties. This is because surface-sensitive methods such as SEM, STM, AFM, etc. provide no input about the inter-grain coupling in the bulk, which is crucial for superconductivity. The macroscopic transport experiments do provide some input through the critical current and critical field measurements. Though these results are difficult to connect directly to the superconducting properties on the local scale.

\subsection{Conclusion.}
To conclude, we revealed the vortex pinning nano-network in thin superconducting films made of sputtered Nb. In the heart of the experiment is a new scanning probe microscopy approach that we named Scanning Quantum Vortex Microscopy (SQVM). The method is based on creating, attracting and dragging a single quantum vortex by the tip of a magnetic force microscope (MFM). The interaction of the moving vortex with defects present in the sample leads to an additional location-dependent pinning force and related dissipation that both modify the characteristics of the MFM cantilever. These characteristics -- oscillation amplitude and phase -- are measured in the experiment and are presented as maps. Since the vortex pinches through the total thickness of a superconductor, it probes both the surface and the bulk components of the pinning potential; this potential is reflected in phase shift maps which represent its surface-projected spatial evolution. We demonstrated the SQVM to enable visualization of defects in superconducting films with a nm resolution over a large field of view. This opens unparalleled possibilities for detailed non-destructive studies of defects inside superconductors and superconducting nano-devices.

\section{Methods}


\textbf{Sample preparation.}
Nb films were fabricated using 2-inch UHV magnetron sputtering. 50, 100 and 240 nm Nb film was deposited onto SiO$_2$(270 nm)/Si substrate. Deposition parameters: pre-etching in Ar plasma $t_{etch}=$~180~sec, $p^{Ar}=2\cdot10^{-2}$ mbar, $P_{RF}=$80 W at $V_{dc}=$580~V, deposition $P_{RF}=$200 W at $V_{dc}$=238~V, $p^{Ar}=4\cdot10^{-3}$ mbar $v_{Nb}=$~0.22~nm/sec. 

\textbf{Transport measurements}.
Electron transport measurements were conducted using a standard four-probe setup in the system Attocube Attodry 1000/SU.

\textbf{High resolution AFM}
The high-resolution topography images were acquired under ambient conditions using the PeakForce method on the Bruker Multimode V8 Atomic Force Microscope. The system utilized an Ostek HA$\_$CNC cantilever with a stiffness of k = 1.5 N/m.


\textbf{Cryogenic MFM}.
The experiments were performed on the AttoCube scanning probe system (AttoDry 1000/SU) utilizing a 9 T superconducting magnet at temperatures between 4K and 10K. The local magnetic field was generated by a magnetic cantilever. The magnetic force microscopy (MFM) measurements utilized a regular magnetic Co/Cr-coated cantilever (MESP, Bruker, with a spring constant of 2.8 N/m). Prior to the measurements, the probes were magnetized at a temperature of 30K and a magnetic field strength of 2,000 Oe. All MFM images were captured in a helium buffer gas environment (at an ambient pressure of 0.5 mbar) and at temperatures ranging from 4 to 30 K, with stabilization maintained within $\pm$0.3 mK. The magnetic contrast was observed in the non-contact lift mode using the phase signal with simultaneous detection of the amplitude signal.

\textbf{SQVM experiment}.
In this work, we demonstrate the spatial and temperature evolution of vortex pinning centers in magnetron-sputtered Nb films with thicknesses of 50, 100, and 240 nm, widely used in superconducting digital and quantum electronics~\cite{fietz1969hysteresis,park1992vortex,dasgupta1978flux}. We observe vortex pinning in niobium films occurring at the boundaries of clustered nanograins, forming a spatially distributed pinning network. We use a standard CoCr-coated magnetic cantilever of a low-temperature magnetic force microscope for the study. At the first stage, applying the field cooling technique, a magnetic cantilever is used to locally generate a single quantum vortex in the film under study; see our recent works~\cite{Dremov_2019,grebenchuk2020observation,Hovhannisyan_2021,stolyarov2022revealing}. The created vortex remains captured by the tip of the cantilever. In the second stage, when the cantilever scans the area of the film under study in a specific temperature range, the vortex is depinned by the tip. It starts examining the superconductor, moving from one pinning center to another. These movements are detected by measuring force gradient and dissipation at the vortex-cantilever interaction~\cite{Auslander_2009}. Since in a type II superconductor, the vortex is clamped across its entire thickness, minimizing its length, our method studies precisely the volumetric pinning potential and obtains its spatial distribution. The spatial resolution of this non-destructive scanning vortex core microscopy (SVCM) is limited only by the superconducting coherence length. We also highlight the possibility of using SVCM to study niobium-based superconducting digital and quantum electronic devices.

\textbf{Modelling and simulations.}
To illustrate the experimental observations, we performed
numerical calculations of the free energy $E$ of the model system
including a point magnetic charge $Q$ as a cantilever located at
height $h$ above a thin superconducting film of thickness $d$. The
film contains an edge high$-j_c$ Josephson junction as a planar
defect positioned in the ($y\,z$) plane. The Josephson junction is
assumed to be described by an ordinary sinusoidal current--phase
relation $j = j_c\,\sin\varphi$. Here $\lambda^d(0)$ and $j_c$ are
the magnetic field penetration depth and Josephson critical
current at zero temperature, respectively. Effective penetration
depth $\lambda^d(T)$ at temperature $T$ is assumed to depend on
the film thickness $d$ as $\lambda^d(T) = \lambda^d(0) / \sqrt{1 -
T^4 / T_c^4(d)}$.  The free energy of the system $E$ consists of
the energy of the total stray magnetic field outside the film, the
kinetic energy of supercurrents created by the point magnetic
charge $Q$ and a vortex in the presence of the defect, and the
energy of Josephson coupling. Figure~\ref{Fig.1}(\textbf{e}) shows
a typical two-dimensional plot of the free energy $E(x_0,\,y_0)$
(\ref{eq:Energy}) as a function of a vortex position ($x_0,\,y_0$)
at low temperature $T \ll T_c$. Two local minima of the free
energy marked as ($A$) and ($B$) in Fig.~\ref{Fig.1}(\textbf{e})
correspond to the positions of the vortex placed under the
cantilever ($x_0 = y_0 = 0$) and a vortex trapped by the defect
($x_0 = t$, $y_0 = 0$). In the second case ($B$) the vortex
structure is more close to the one for a Josephson vortex.
Figure~\ref{Fig.1}(\textbf{f}) shows the dependence of the free
energy $E$ on temperature $T$ for ($A$) and ($B$) arrangements of
a vortex (see panel (\textbf{e})) for a fixed place ($h,\,t$) of
the magnetic charge $Q$ (cantilever) with respect to the planar
defect, and for several values of the penetration depth
$\lambda^d(0)$. We put $\lambda^d(0) =
\mathrm{(130;\,100;\,80)\,nm}$ for films $d =
\mathrm{(50;\,100;\,240)\,nm}$, respectively \cite{Gubin}. One can
observe the shift of the global minimum of the energy
($\mathrm{min}\{E_A,\,E_B\}$) from the position at the defect
($B$) to the position under the cantilever ($A$) with increasing
temperature $T$. The crossover temperature $T^*$ between the two
regimes grows with the reduction of the penetration depth
$\lambda^d(0)$, which corresponds to the increase in the film
thickness $d$ \cite{Gubin}. These calculations clearly correlate
with the observations of reduction of the temperature range
$\Delta T = T_c - T^*$ where defects can be visualized, as the
film thickness increases (see
Fig.~\ref{Fig.2}(\textbf{e}),(\textbf{h})). The plots
Figs.~\ref{Fig.3}(\textbf{d}),(\textbf{h}) illustrate a strongly
different behavior of the vortex energy during the scanning
procedure. At low temperatures $T \ll T_c$ the shift of the probe
with respect to the defect does not change the relation between
two local energy minima: the global minimum of the free energy
always corresponds to the vortex trapped by the defect
(Fig.~\ref{Fig.3}(\textbf{d})). For temperatures $T$ close to
$T_c$ the relation between two local energy minima  changes as the
cantilever shifts: at a sufficiently large distance of the
cantilever from the defect the global minimum of the free energy
corresponds to the vortex positioned under the cantilever
(Fig.~\ref{Fig.3}(\textbf{h})).

In order to confirm our explanations of the experimentally
observed features we consider a simple model problem of the vortex
pinning by a planar defect in a thin superconducting (SC) film of
thickness $d$ much less the London penetration depth $\lambda_0$
at zero temperature ($d \ll \lambda_0$). In this limit the
screening properties of the film are determined by the Pearl
length $\Lambda = \lambda^2 / d$~\cite{Pearl_1964,Pearl_1966}. To
model the vortex pinning we introduce a high$-j_c$ planar defect
described by a standard Josephson current--phase relation $j =
j_c\,\sin\varphi$. A MFM cantilever is modelled by a point
magnetic charge $Q$ which is positioned at a height $h$ above the
film and is shifted at the distance $t$ in the film plane with
respect to the defect. In the absence of superconducting film the
magnetic field of the cantilever $\mathbf{B}^Q = \mathrm{curl}
\mathbf{A}^Q$ is described by the vector potential $\mathbf{A}^Q =
A^Q(r,\,z)\,\mathbf{\theta}_0$
\begin{equation}\label{eq:AQ}
    A^Q(r,\,z) = \frac{Q}{4 \pi r} \left( 1 +  \frac{z - h}{\sqrt{r^2 + (z-h)^2}}
    \right)\, ,
\end{equation}
where ($r,\,\theta,\,z$) the polar coordinate system.
We assume here that the induced current density
$\mathbf{j}(\mathbf{r})$ is so weak that the suppression of the
magnitude of the superconducting order parameter is negligible
everywhere in the film except for the vortex core. The
London equation for the sheet current
$\mathrm{\mathbf{g}}(\mathbf{r}) = \mathbf{j}(\mathbf{r})\,d$
after averaging over the film thickness reads
\begin{equation}\label{eq:SC}
    \mathrm{\mathbf{g}}(\mathbf{r}) = \frac{c}{4 \pi \Lambda} \left(\,
    \mathbf{S} - \mathbf{A} - \mathbf{A}^Q\, \right)\,\delta(z),
\end{equation}
where $\mathbf{B} = \mathrm{curl} \mathbf{A}$ is the magnetic
field of the current $\mathrm{\mathbf{g}}(\mathbf{r}) =
(\mathrm{g_x},\,\mathrm{g_y})$
\begin{equation}\label{eq:L-M}
    \mathrm{curl}\,\mathrm{curl}\,\mathbf{A} =
        \frac{4\,\pi}{c}\,\mathrm{\mathbf{g}}(\mathbf{r})\,,
\end{equation}
and the term $\mathbf{S} = \mathbf{S}^P + \mathbf{S}^D$ describes
the vortex source at the point $\mathbf{r}_0 =
(x_0,\,y_0)$ \cite{Pearl_1964}
\begin{equation}\label{eq:SP}
    \mathrm{curl}\, \mathbf{S}^P = \Phi_0\, \delta(\mathbf{r}-\mathbf{r}_0)\, \mathbf{z}_0\,,
\end{equation}
and the planar defect at the plane $x = t$ with the phase
difference $\varphi(y)$
\cite{Ivanchenko-pla90,Mints-prb94,Mints-prb95,Kogan-prb01}
\begin{equation}\label{eq:SD}
    \mathrm{curl}\, \mathbf{S}^D = \frac{\Phi_0}{2\,\pi}\, \frac{d \varphi}{d
    y}\,\delta(x-t)\,\mathbf{z}_0\,,
\end{equation}
$\mathbf{r} = (x,\,y)$, and $\Phi_0 = \pi \hbar\, c / e$ is the
flux quantum. For an arbitrary position ($x_0,\,y_0$) of the
vortex with respect to the cantilever and the planar defect and
for a given phase difference $\varphi(y)$, the linear
equations~(\ref{eq:AQ})--(\ref{eq:SD}) can be solved by standard
Fourier method.

To obtain a self-consistent equation for the phase difference
$\varphi(y)$ we use the continuity condition for the normal
component of the sheet current $\mathrm{g}_x(t,\,y)$ flowing
through the planar defect
\begin{equation}\label{eq:BC}
    \mathrm{g}_x(t,\,y) = \mathrm{g}_c\,\sin\,\varphi(y)\,,
\end{equation}
where $\mathrm{g}_c = j_c d$ is the critical current of the edge
junction in the thin film. The magnetic field of the cantilever
$\mathbf{B}^Q$ and the Abrikosov vortex trapped in one of the
banks of an edge-type planar junction cause an extra phase
difference on the junction that depends on both an inhomogeneous
magnetic field of the probe
\cite{Samokhvalov-jetpl03,Dremov_2019} and the vortex
positions with respect to the defect
\cite{Krasnov-prl10,Clem-prb11,Kogan-Mints-prb14,Mironov-prb17,Krasnov-prb19}.
The Josephson effect at such edge junction is quite different from
those at familiar bulk junctions, because the stray magnetic field
results in an integral equation governing the phase distribution
$\varphi(y)$, i.e., the problem becomes nonlocal
\cite{Ivanchenko-pla90,Mints-prb94,Mints-prb95,Kogan-prb01}. The
edge Josephson junction in the thin film is characterized by two
characteristic lengths. The first one is the Pearl length
$\Lambda$ which describes the magnetic field screening by the SC
film. Another length $L = \lambda_J^2 / \lambda$ (here $\lambda_J
= \sqrt{c \Phi_0 / 16 \pi^2 j_c \lambda}$ is the Josephson length)
characterizes the junction properties. Using the boundary
condition (\ref{eq:BC}) and the solution of the
Eqs.~(\ref{eq:AQ})-(\ref{eq:SD}) one finally obtains the following
integral equation for the phase difference $\varphi(y)$
\begin{gather}
        \Lambda\,\int\limits_{-\infty}^{\infty} d s\, \varphi_s^{\prime\prime}(s)\, G_0 (0,\,y-s)
        = \mu\,\sin\varphi(y) - \frac{4\,Q\,y}{\Phi_0 \sqrt{y^2+t^2}}\,
                \int\limits_0^\infty d q \frac{q\,J_1(q\,\sqrt{y^2+t^2}/\Lambda)\,
                \mathrm{e}^{-q\,h / \Lambda}}{1 + 2  q} +  \label{eq:BasicEq} \\ +
    \frac{\pi\,(y-y_0)}{\sqrt{(y-y_0)^2 +
    (t-x_0)^2}}\,G_1\left(\,t-x_0,\,y-y_0\,\right)\,, \nonumber
\end{gather}
where the dimensionless parameter $\mu = 2 \Lambda / L$ describes the
properties of the planar defect, and
$$
        G_0(u,\,v) =  H_0\left( \frac{\sqrt{u^2+v^2}}{2\Lambda} \right)
        - Y_0\left(\frac{\sqrt{u^2+v^2}}{2\Lambda}\right)\,,
\qquad
        G_1(u,\,v) =  H_1\left( \frac{\sqrt{u^2+v^2}}{2\Lambda} \right)
        - Y_1\left(\frac{\sqrt{u^2+v^2}}{2\Lambda}\right) -
        \frac{2}{\pi}\,.
$$
Here, $H_{0,1}$, $Y_{0,1}$ are Struve and second-kind Bessel
functions of the  order 0 an 1 (see
Ref.~\cite{Abramowitz-Stegun}). The equation (\ref{eq:BasicEq})
was solved numerically using the iteration method on a discrete
grid of nodes with a step $\Delta y = \mathrm{0.005} \Lambda$ and
the accuracy better than $\mathrm{10^{-5}}$. The total energy $E =
E_J + E_\mathrm{\mathbf{g}} + E_\mathbf{B}$ of the system under
consideration consists of the Josephson coupling energy ($E_J$),
the kinetic energy ($E_\mathrm{\mathbf{g}}$) of the supercurrents,
and the energy ($E_\mathbf{B}$) of stray magnetic fields in the
surrounding space:
\begin{gather}\label{eq:Energy}
        E = \frac{\Phi_0^2}{32 \pi^3 \Lambda L} \int d y \left[\, 1 - \cos\varphi(y)\, \right] +
        \frac{2 \pi \Lambda}{c^2} \int d \mathbf{r}\,\mathrm{\mathbf{g}}^2(\mathbf{r}) +
        \frac{1}{8 \pi} \int d \mathbf{r}\, d z\, \mathbf{B}^2\,,
\end{gather}
We performed numerical calculations of the free energy $E$
(\ref{eq:Energy}) as a function of a vortex position ($x_0,\,y_0$)
for different values of temperature $T$, the lift of the
cantilever $h$ and shift of the cantilever with respect to the
defect $t$ (see panels (\textbf{e}), (\textbf{f}) in
Fig.~\ref{Fig.1} and panels (\textbf{d}), (\textbf{h}) in
Fig.~\ref{Fig.3}).

\
\subsection{Data availability.}
Authors can confirm that all relevant data are included in the paper and its supplementary information files. Additional data are available on request from the authors.


\begin{thebibliography}{0}%
\makeatletter
\providecommand \@ifxundefined [1]{%
 \@ifx{#1\undefined}
}%
\providecommand \@ifnum [1]{%
 \ifnum #1\expandafter \@firstoftwo
 \else \expandafter \@secondoftwo
 \fi
}%
\providecommand \@ifx [1]{%
 \ifx #1\expandafter \@firstoftwo
 \else \expandafter \@secondoftwo
 \fi
}%
\providecommand \natexlab [1]{#1}%
\providecommand \enquote  [1]{``#1''}%
\providecommand \bibnamefont  [1]{#1}%
\providecommand \bibfnamefont [1]{#1}%
\providecommand \citenamefont [1]{#1}%
\providecommand \href@noop [0]{\@secondoftwo}%
\providecommand \href [0]{\begingroup \@sanitize@url \@href}%
\providecommand \@href[1]{\@@startlink{#1}\@@href}%
\providecommand \@@href[1]{\endgroup#1\@@endlink}%
\providecommand \@sanitize@url [0]{\catcode `\\12\catcode `\$12\catcode
  `\&12\catcode `\#12\catcode `\^12\catcode `\_12\catcode `\%12\relax}%
\providecommand \@@startlink[1]{}%
\providecommand \@@endlink[0]{}%
\providecommand \url  [0]{\begingroup\@sanitize@url \@url }%
\providecommand \@url [1]{\endgroup\@href {#1}{\urlprefix }}%
\providecommand \urlprefix  [0]{URL }%
\providecommand \Eprint [0]{\href }%
\providecommand \doibase [0]{https://doi.org/}%
\providecommand \selectlanguage [0]{\@gobble}%
\providecommand \bibinfo  [0]{\@secondoftwo}%
\providecommand \bibfield  [0]{\@secondoftwo}%
\providecommand \translation [1]{[#1]}%
\providecommand \BibitemOpen [0]{}%
\providecommand \bibitemStop [0]{}%
\providecommand \bibitemNoStop [0]{.\EOS\space}%
\providecommand \EOS [0]{\spacefactor3000\relax}%
\providecommand \BibitemShut  [1]{\csname bibitem#1\endcsname}%
\let\auto@bib@innerbib\@empty
\end{thebibliography}%


\begin{thebibliography}{99}

 \bibitem{menard2015coherent} G. C. M'enard, S. Guissart, C. Brun, S. Pons, V. S. Stolyarov, F. Debontridder, M. V. Leclerc, E. Janod, L. Cario, D. Roditchev, et al., Coherent long-range magnetic bound states in a superconductor.{\em Nature Physics} \textbf{11}, 1013 (2015)

\bibitem{stolyarov2018domain}V. S. Stolyarov, I. S. Veshchunov, S. Y. Grebenchuk, D. S. Baranov, I. A. Golovchanskiy, A. G. Shishkin, N. Zhou, Z. Shi, X. Xu, S. Pyon, et al., Domain Meissner state and spontaneous vortex-antivortex generation in the ferromagnetic superconductor. {\em Science advances} \textbf{4}, eaat1061 (2018).

\bibitem{grebenchuk2020crossover} S. Y. Grebenchuk, Z. A. Devizorova, I. Golovchanskiy, I. Shchetinin, G.-H. Cao, A. Buzdin, D. Roditchev, and V. Stolyarov, Crossover from ferromagnetic superconductor to superconducting ferromagnet in P-doped EuFe$_2$ (As$_{1-x}$ P$_x$)$_2$. {\em Physical Review B} \textbf{102}, 144501 (2020).

\bibitem{stolyarov2022effective} V. Stolyarov, V. Oboznov, D. Kasatonov, A. Neilo, S. Bakurskiy, N. Klenov, I. Soloviev, M. Kupriyanov, A. Golubov, T. Cren, et al., Effective Exchange Energy in a Thin, Spatially Inhomogeneous CuNi Layer Proximized by Nb. {\em The Journal of Physical Chemistry Letters} \textbf{13}, 6400 (2022).

\bibitem{vagov2023intertype} A. Vagov, T. T. Saraiva, A. A. Shanenko, A. S. Vasenko, J. A. Aguiar, V. S. Stolyarov, and D. Roditchev,  Intertype superconductivity in ferromagnetic superconductors. {\em Communications Physics} \textbf{6}, 284 (2023).

\bibitem{Larbalestier_2018} Z.-H. Sung, P. J. Lee, A. Gurevich, and D. C. Larbalestier, Evidence for preferential flux flow at the grain boundaries of superconducting RF-quality niobium, \textit{ Supercond. Sc. Technol.} {\bf 31}, 045001 (2018).

\bibitem{golod2021reconfigurable} T. Golod, R. A. Hovhannisyan, O. M. Kapran, V. V. Dremov, V. S. Stolyarov, and V. M. Krasnov, Reconfigurable Josephson phase shifter. {\em Nano Letters} \textbf{21}, 5240 (2021).

\bibitem{grebenchuk2020observation} S. Y. Grebenchuk, R. A. Hovhannisyan, V. V. Dremov, A. G. Shishkin, V. I. Chichkov, A. A. Golubov, D. Roditchev, V. M. Krasnov, and V. S. Stolyarov, Observation of interacting Josephson vortex chains by magnetic force microscopy. {\em Physical Review Research} \textbf{2}, 023105 (2020).

\bibitem{grebenchuk2022magnetic} S. Y. Grebenchuk, R. Hovhannisyan, A. Shishkin, V. Dremov, and V. Stolyarov, Magnetic Force Microscopy for Diagnosis of Complex Superconducting Circuits. {\em Physical Review Applied} \textbf{18}, 054035 (2022).

\bibitem{golod2022demonstration} T. Golod and V. M. Krasnov, Demonstration of a superconducting diode-with-memory, operational at zero magnetic field with switchable nonreciprocity. {\em Nature Communications} \textbf{13}, 3658 (2022).

\bibitem{Golovchanskiy_JAP_114_163910} I. A. Golovchanskiy, A. V. Pan, O. V. Shcherbakova, and S. A. Fedoseev, Rectifying differences in transport, dynamic, and quasi-equilibrium
measurements of critical current density, {\em J. Appl. Phys.}, \textbf{114}, 163910 (2013).

\bibitem{Golovchanskiy_ASS_311_549} Igor A. Golovchanskiy, Alexey V. Pan, Sergey A. Fedoseev, Michael Higgins, Significant tunability of thin film functionalities enabled by manipulating magnetic and structural nano-domains, {\em Applied Surface Science} \textbf{311}, 549 (2014).



\bibitem{Annunziata_2010} Anthony J Annunziata et al Tunable superconducting nanoinductors, \textit{Nanotechnology} \textbf{21} 445202 (2010)

\bibitem{carbillet2020spectroscopic} C. Carbillet, V. Cherkez, M. Skvortsov, M. Feigel'man, F. Debontridder, L. Ioffe, V. Stolyarov, K. Ilin, M. Siegel, D. Roditchev, et al., Spectroscopic evidence for strong correlations between local superconducting gap and local Altshuler-Aronov density of states suppression in ultrathin NbN films. {\em Physical Review B} \textbf{102}, 024504 (2020).


\bibitem{Soloviev_2021} I. I. Soloviev, V. I. Ruzhickiy, S. V. Bakurskiy, N. V. Klenov, M. Yu. Kupriyanov, A. A. Golubov, O. V. Skryabina, and V. S. Stolyarov, Superconducting Circuits without Inductors Based on Bistable Josephson Junctions, Phys. Rev. Applied 16, 014052 (2021)

\bibitem{Campbell1972} A.~Campbell and J.~E.~Evetts, \textit{Critical Currents in Superconductors} (Taylor and Francis, London, 1972).
  
  \bibitem{Blatter1994} G.~Blatter, M.~V.~Feigel'man, V.~B.~Geshkenbein, A.~I.~Larkin, and V.~M.~Vinokur, Vortices in high-temperature superconductors, Rev. Mod. Phys. \textbf{66}, 1125 (1994).

  \bibitem{BezryadinPLA1994} A.~Bezryadin, A.~Buzdin, B.~Pannetier, Phase transitions in a superconducting thin film with a single circular hole, Phys. Lett. A \textbf{195}, 373 (1994).
	
  \bibitem{BezryadinJLTP1995} A.~Bezryadin and B.~Pannetier, Nucleation of Superconductivity in a Thin Film with a Lattice of Circular Holes, J. Low Temp. Phys. \textbf{98}, 251 (1995).
	
  \bibitem{BerdiyorovEPL2006} G.~R.~Berdiyorov, M.~V.~Milo\v{s}evi\'{c}, and F.~M.~Peeters, Superconducting films with antidot arrays - Novel behavior of the critical current, Europhys. Lett. \textbf{74}, 493 (2006).
	
  \bibitem{BerdiyorovPRL2006} G.~R.~Berdiyorov, M.~V.~Milo\v{s}evi\'{c}, and F.~M.~Peeters, Novel Commensurability Effects in Superconducting Films with Antidot Arrays, Phys. Rev. Lett. \textbf{96}, 207001 (2006).
	
  \bibitem{BerdiyorovPRB2006} G.~R.~Berdiyorov, M.~V.~Milo\v{s}evi\'{c}, and F.~M.~Peeters, Vortex configurations and critical parameters in superconducting thin films containing antidot arrays: Nonlinear Ginzburg-Landau theory, Phys. Rev. B \textbf{74}, 174512 (2006).
	
  \bibitem{SabatinoJAP2010} P.~Sabatino, C.~Cirillo, G.~Carapella, M.~Trezza, and C.~Attanasio, High field vortex matching effects in superconducting Nb thin films with a nanometer-sized square array of antidots, J. Appl. Phys. \textbf{108}, 053906 (2010).
	
  \bibitem{LatimerPRB2012} M.~L.~Latimer, G.~R.~Berdiyorov, Z.~L.~Xiao, W.~K.~Kwok, and F.~M.~Peeters, Vortex interaction enhanced saturation number and caging effect in a superconducting film with a honeycomb array of nanoscale holes, Phys. Rev. B \textbf{85}, 012505 (2012).
	
  \bibitem{LatimerPRL2013} M.~L.~Latimer, G.~R.~Berdiyorov, Z.~L.~Xiao, F.~M.~Peeters, and W.~K.~Kwok, Realization of Artificial Ice Systems for Magnetic Vortices in a Superconducting MoGe Thin Film with Patterned Nanostructures, Phys. Rev. Lett. \textbf{111}, 067001 (2013).
	
  \bibitem{GePRB2017} J.-Y.~Ge, V.~N~Gladilin, J.~Tempere, V.~S.~Zharinov, J.~Van de Vondel, J.~T.~Devreese, and V.~V.~Moshchalkov, Direct visualization of vortex ice in a nanostructured superconductor, Phys. Rev. B \textbf{96}, 134515 (2017).
	
  \bibitem{XueNJP2018} C.~Xue, J.-Y.~Ge, A.~He, V.~S.~Zharinov, V.~V.~Moshchalkov, and Y.-H.~Zhou, Stability of degenerate vortex states and multi-quanta confinement effects in a nanostructured superconductor with Kagome lattice of elongated antidots, New J. Phys. \textbf{20}, 093030 (2018).
  \bibitem{BezryadinPRB1996} A.~Bezryadin, Yu.~N.~Ovchinnikov, and B.~Pannetier, Nucleation of vortices inside open and blind microholes, Phys. Rev. B \textbf{53}, 8553 (1996).
	
  \bibitem{BerdiyorovNJP2009} G.~R.~Berdiyorov, M.~V.~Milo\v{s}evi\'{c}, and F.~M.~Peeters, Composite vortex ordering in superconducting films with arrays of blind holes, New J. Phys. \textbf{11}, 013025 (2009).
  \bibitem{KarapetrovPRL2005} G.~Karapetrov, J.~Fedor, M.~Iavarone, D.~Rosenmann, and W.~K.~Kwok, Direct Observation of Geometrical Phase Transitions in Mesoscopic Superconductors by Scanning Tunneling Microscopy, Phys. Rev. Lett. \textbf{95}, 167002 (2005).
	
  \bibitem{SadovskyyPRB2017} I.~A.~Sadovskyy, Y.~L.~Wang, Z.-L.~Xiao, W.-K.~Kwok, and A.~Glatz, Effect of hexagonal patterned arrays and defect geometry on the critical current of superconducting films, Phys. Rev. B \textbf{95}, 075303 (2017).
	
  \bibitem{WillaSUST2018} R.~Willa, A.~E.~Koshelev, I.~A.~Sadovskyy, and A.~Glatz, Strong-pinning regimes by spherical inclusions in anisotropic type-II superconductors, Supercond. Sci. Technol. \textbf{31}, 014001 (2018).
  \bibitem{KwokRPP2016} W.-K.~Kwok, U.~Welp, A.~Glatz, A.~E.~Koshelev, K.~J.~Kihlstrom, and G.~W.~Crabtree, Vortices in high-performance high-temperature superconductors, Rep. Prog. Phys. \textbf{79}, 116501 (2016).
  \bibitem{BeanPRL1971} C.~P.~Bean and J.~D.~Livingston, Surface Barrier in Type-II Superconductors, Phys. Rev. Lett. \textbf{12}, 14 (1964).

  \bibitem{DaldiniPRL1974} O.~Daldini, P.~Martinoli, J.~L.~Olsen, and G.~Berner, Vortex-Line Pinning by Thickness Modulation of Superconducting Films, Phys. Rev. Lett. \textbf{32}, 218 (1974).
  \bibitem{CrisanAPL2001} A.~Crisan, S.~Fujiwara, J.~C.~Nie, A.~Sundaresan, and H.~Ihara, Sputtered nanorods: A costless method for inducing effective pinning centers in superconducting thin films, Appl. Phys. Lett. \textbf{79}, 4547 (2001).
  \bibitem{MartinPRL1997} J.~I.~Mart\'{i}n, M.~V\'{e}lez, J.~Nogu\'{e}s, and I.~K.~Schuller, Flux Pinning in a Superconductors by an Array of Submicrometer Magnetic Dots, Phys. Rev. Lett. \textbf{79}, 1929 (1997).
	
  \bibitem{VanBaelPRB1999} M.~J.~Van Bael, K.~Temst, V.~V.~Moshchalkov, and Y.~Bruynseraede, Phys. Rev. B \textbf{59}, 14674 (1999).
  \bibitem{FeighanSUST2017} J.~P.~F.~Feighan, A.~Kursumovic, and J.~L.~MacManus-Driscoll, Materials design for artificial pinning centres in superconductor PLD coated conductors, Supercond. Sci. Technol. \textbf{30}, 123001 (2017).

  \bibitem{BugoslavskyN2001} Y.~Bugoslavsky, L.~F.~Cohen, G.~K.~Perkins, M.~Polichetti, T.~J.~Tate, R.~Gwilliam, and A.~D.~Caplin, Enhancement of the high-magnetic-field critical density of superconducting MgB$_2$ by proton irradiation, Nature \textbf{411}, 561 (2001).
	
  \bibitem{NakajimaPRB2009} Y.~Nakajima, Y.~Tsuchiya, T.~Taen, T.~Tamegai, S.~Okayasu, and M.~Sasase, Enhancement of critical current density in Co-doped BaFe$_2$As$_2$ with columnar defects introduced by heavy-ion irradiation, Phys. Rev. B \textbf{80}, 012510 (2009).
	
  \bibitem{ZechnerSUST2018} G.~Zechner, K.~L.~Mletschnig, W.~Lang, M.~Dosmailov, M.~A.~Bodea, and J.~D.~Pedarnig, Unconventional critical state in YBa$_{2}$Cu$_{3}$O$_{7-\delta}$ thin films with a vortex-pin lattice fabricated by masked He$^+$ ion beam irradiation, Supercond. Sci. Technol. \textbf{31}, 044002 (2018).
	
  \bibitem{AntonovPSS2019} A.~V.~Antonov, A.~V.~Ikonnikov, D.~V.~Masterov, A.~N.~Mikhaylov, S.~V.~Morozov, Yu.~N.~Nozdrin, S.~A.~Pavlov, A.~E.~Parafin, D.~I.~Tetel'baum, S.~S.~Ustavschikov, P.~A.~Yunin, and D.~A.~Savinov, Phase Diagrams of Thin Disordered Films Based on HTSC YBa$_{2}$Cu$_{3}$O$_{7-x}$ in External Magnetic Fields, Phys. Solid State \textbf{61}, 1523 (2019).
	
  \bibitem{AntonovPSS2020} A.~V.~Antonov, A.~I.~El'kina, V.~K.~Vasiliev, M.~A.~Galin, D.~V.~Masterov, A.~N.~Mikhaylov, S.~V.~Morozov, S.~A.~Pavlov, A.~E.~Parafin, D.~I.~Tetelbaum, S.~S.~Ustavschikov, P.~A.~Yunin, and D.~A.~Savinov, Experimental Observation of s-Component of Superconducting Pairing in Thin Disordered HTSC Films Based on YBCO, Phys. Solid State \textbf{62}, 1598 (2020).
	
  \bibitem{AntonovPC2020} A.~V.~Antonov, A.~V.~Ikonnikov, D.~V.~Masterov, A.~N.~Mikhaylov, S.~V.~Morozov, Yu.~N.~Nozdrin, S.~A.~Pavlov, A.~E.~Parafin, D.~I.~Tetel'baum, S.~S.~Ustavschikov, V.~K.~Vasiliev, P.~A.~Yunin, D.~A.~Savinov, Critical-field slope reduction and upward curvature of the phase-transition lines of thin disordered superconducting YBa$_2$Cu$_{3}$O$_{7-x}$ films in strong magnetic fields, Phys. C: Supercond. Appl. \textbf{568}, 1353581 (2020).
	
  \bibitem{AichnerFNT2020} B.~Aichner, K.~L.~Mletschnig, B.~M\"{u}ller, M.~Karrer, M.~Dosmailov, J.~D.~Pedarnig, R.~Kleiner, D.~Koelle, and W.~Lang, Angular magnetic-field dependence of vortex matching in pinning lattices fabricated by focused or masked helium ion beam irradiation of superconducting YBa$_2$Cu$_3$O$_{7-\delta}$ thin films, Fiz. Nizk. Temp. \textbf{46}, 402 (2020).





\bibitem{Zheng_2015} L. J. Zeng, S. Nik, T. Greibe, P. Krantz, C. M. Wilson, P. Delsing, and E. Olsson, Direct observation of the thickness distribution of ultra thin AlOx barriers in Al/AlO$_x$/Al Josephson junctions, \textit{ J. Phys. D: Appl. Phys.} {\bf 48}, 395308 (2015).

\bibitem{kudriashov2022revealing} A. Kudriashov, I. Babich, R. A. Hovhannisyan, A. G. Shishkin, S. N. Kozlov, A. Fedorov, D. V. Vyalikh, E. Khestanova, M. Y. Kupriyanov, and V. S. Stolyarov, Revealing Intrinsic Superconductivity of the Nb/BiSbTe2Se Interface {\em Advanced Functional Materials} \textbf{32}, 2209853 (2022).

\bibitem{golovchanskiy2023magnetization} I. Golovchanskiy, N. Abramov, O. Emelyanova, I. Shchetinin, V. Ryazanov, A. Golubov, and V. Stolyarov, Magnetization Dynamics in Proximity-Coupled Superconductor-Ferromagnet-Superconductor Multilayers. II. Thickness Dependence of the Superconducting Torque. {\em Physical Review Applied} {\bf 19}, 034025 (2023).

\bibitem{Devred_2004} A. Devred, Practical Low-Temperature Superconductors for Electromagnets, \textit{CERN Reports} {\bf 4} (2004), ISSN 0007-8328, DOI:10.5170/CERN-2004-006.

\bibitem{martinez2013microstructures} E. Martinez, L. E. Murr, J. Hernandez, X. Pan, K. Amato, P. Frigola, C. Terrazas, S. Gaytan, E. Rodriguez, F. Medina, et al., Microstructures of niobium components fabricated by electron beam melting.  {\em Metallography, microstructure, and analysis} {\bf 2}, 183 (2013).

\bibitem{stolyarov2018expansion} V. S. Stolyarov, T. Cren, C. Brun, I. A. Golovchanskiy, O. V. Skryabina, D. I. Kasatonov, M. M. Khapaev, M. Y. Kupriyanov, A. A. Golubov, and D. Roditchev, Expansion of a superconducting vortex core into a diffusive metal. {\em Nature communications} \textbf{9}, 2277 (2018).

\bibitem{berti2023scanning} G. Berti, C. Torres-Castanedo, D. Goronzy, M. Bedzyk, M. Hersam, C. Kopas, J. Marshall, and M. Iavarone, Scanning tunneling microscopy and spectroscopy characterization of Nb films for quantum applications. {\em Applied Physics Letters}  \textbf{122} (2023).

\bibitem{koslowski2004studying} B. Koslowski, C. Dietrich, and P. Ziemann, Studying the bulk band structure of niobium by scanning tunneling spectroscopy. {\em Surface science} \textbf{557}, 255 (2004).

\bibitem{Lee_2011} J. Lee, H. Wang, M. Dreyer, H. Berger, and B. I. Barker, Nonuniform and coherent motion of superconducting vortices in the picometer-per-second regime, \textit{ Phys. Rev. B}{\bf 84}, 060515(R) (2011).

\bibitem{wu2005studies} G. Wu, A.-M. Valente, H. Phillips, H. Wang, A. Wu, T. Renk, and P. Provencio, Studies of niobium thin film produced by energetic vacuum deposition. {\em Thin Solid Films} \textbf{489}, 56 (2005).

 

\bibitem{Vlasko_1992} L.A. Dorosinskii, M.V. Indenbom, V.I. Nikitenko, Yu.A.Ossip'yan, A.A. Polyanskii, V.K. Vlasko-Vlasov, Studies of HTSC crystal magnetization features using indicator magneto-optic films with in-plane anisotropy, \textit{ Physica C} {\bf 203} 149 (1992)

\bibitem{Veshchunov_2016} I.S. Veshchunov, W. Magrini, S.V. Mironov, A.G. Godin, J.-B. Trebbia, A.I. Buzdin, Ph. Tamarat, and B. Lounis. Optical manipulation of single flux
quanta. \textit{Nature Commun.} {\bf 7}, 12801 (2016).

\bibitem{Koblischka_1995} M. R. Koblischka and R. J. Wijngaarden, Magneto-optical investigations of superconductors, \textit{ Supercond. Sci. Technol.} {\bf 8}, 199 (1995).

\bibitem{Runge_2000} B.-U. Runge, U. Bolz, J. Boneberg, V. Bujok, P. Br\"{u}ll, J. Eisenmenger,J. Schiessling, and P. Leiderer, Magneto-Optic Characterization of Defects and Study of Flux Avalanches in High-$T_c$  Superconductors down to Nanosecond Time Resolution, \textit{Laser Phys.} {\bf 10}, 53 (2000).




\bibitem{Tonomura_1992} K. Harada, T. Matsuda, J. Bonevich, M. Igarashi, S. Kondo, G. Pozzi, U. Kawabet, and A. Tonomura, Real-time observation of vortex lattices in a superconductor by electron microscopy.\textit{ Nature} {\bf 360}, 51 (1992).

\bibitem{Vinnikov_2012} L.Ya. Vinnikov, A. G. Troshina, I. S. Veschunov, D. Analytis, I. R. Fisher, Yu. Liu, C. T. Lin, L. Fang, U. Welp, and W. K. Kwok, Vortex Structure in BaFe$_2$(As$_{1 \cdot x}$P$_x$)$_2$ Single Crystals, \textit{ JETP Lett.} {\bf 96}, 655 (2012).

\bibitem{Vinnikov_2019} L.Y. Vinnikov, I.S. Veshchunov, M.S. Sidelnikov, V.S. Stolyarov, The High-Resolution Bitter Decoration Technique for the Magnetic Flux Structure Imaging at Low Temperatures, \textit{Instrum Exp Tech} \textbf{62}, 587-593 (2019).

\bibitem{halbertal2016nanoscale} D. Halbertal, J. Cuppens, M. B. Shalom, L. Embon, N. Shadmi, Y. Anahory, H. Naren, J. Sarkar, A. Uri, Y. Ronen, et al., Nanoscale thermal imaging of dissipation in quantum systems. {\em  Nature} {\bf 539}, 407 (2016).

\bibitem{finkler2012nano}A. Finkler, D. Vasyukov, Y. Segev, L. Neeman, Y. Anahory, Y. Myasoedov, M. Rappaport, M. Huber, J. Martin, A. Yacoby, et al., Nano-sized SQUID-on-tip for scanning probe microscopy {\em Journal of Physics: Conference Series},  {\bf 400}, 052004 (2012).


\bibitem{Embon_2017} L. Embon, Y. Anahory, \v{Z}.L. Jeli\'{c}, E.O. Lachman, Y. Myasoedov, M.E. Huber, G.P. Mikitik, A.V. Silhanek, M.V. Milo\v{s}evi\'{c}, A. Gurevich, and E. Zeldov, Imaging of super-fast dynamics and flow instabilities of superconducting vortices, \textit{ Nat. Commun.} {\bf 8}, 85 (2017).

\bibitem{Kirtley_2016} J. R. Kirtley, L. Paulius, A. J. Rosenberg, J. C. Palmstrom, C. M. Holland, E. M. Spanton, D. Schiessl, C. L. Jermain, J. Gibbons, Y.-K.-K. Fung, M. E. Huber, D.C. Ralph, M.B. Ketchen, G.W. Gibson and K. A. Moler, Scanning SQUID susceptometers with submicrometer spatial resolution, \textit{ Rev. Sci. Instrum.} {\bf 87},
093702 (2016).

\bibitem{Embon_2015} Embon, L., Anahory, Y., Suhov, A. et al. Probing dynamics and pinning of single vortices in superconductors at nanometer scales. \textit{Sci Rep} \textbf{5}, 7598 (2015). 


\bibitem{Tamegai_2001} A. Grigorenko, S. Bending, T. Tamegai,S. Ooi, and M. Henini, A one-dimensional chain state of vortex matter, \textit{ Nature} {\bf 414}, 728 (2001).

\bibitem{Kalisky_2009} B. Kalisky, J. R. Kirtley, E. A. Nowadnick, R. B. Dinner, E. Zeldov, Ariando, S. Wenderich, H. Hilgenkamp, D. M. Feldmann, and K. A. Moler, Dynamics of single vortices in grain boundaries: I-V characteristics on the femtovolt scale, \textit{ Appl. Phys. Lett.} {\bf 94}, 202504 (2009).

\bibitem{Park_2012} S.K. Park, B.R. Cho, H.Y. Park, H.-C. Ri, Analysis of the local current in GdBCO coated conductors using low-temperature scanning laser and Hall probe microscopy, \textit{ Cryogenics} {\bf 52}, 744 (2012).

\bibitem{Moschalkov_2015} J.-Y. Ge, J. Gutierrez, V. N. Gladilin, J. T. Devreese, and V. V. Moshchalkov, Bound vortex dipoles generated at pinning centres by Meissner current, \textit{ Nature Commun.} {\bf 6}, 6573 (2015).

\bibitem{Volodin_2002} A. Volodin, K. Temst, C. Van Haesendonck, Y. Bruynseraede, M. I. Montero and I. K. Schuller, Magnetic-force microscopy of vortices in thin niobium films: Correlation between the vortex distribution and the thickness-dependent film morphology, \textit{ Europhys. Lett.} {\bf 58}, 582 (2002).

\bibitem{Auslander_2009} O. M. Auslaender, L. Luan, E.W.J. Straver, J. E. Hoffman, N. C. Koshnick, E. Zeldov, D. A. Bonn, R. Liang, W. N. Hardy and K. A. Moler, Mechanics of individual isolated vortices in a cuprate superconductor, \textit{ Nature Phys. } {\bf 5}, 35 (2009).

\bibitem{Budakian_2019} H. Polshyn, T. R. Naibert, and R. Budakian, Manipulating multivortex states in superconducting structures, \textit{ Nano Lett.} {\bf 19}, 5476 (2019).

\bibitem{Dremov_2019} V. V. Dremov, S. Y. Grebenchuk, A. G. Shishkin, D. S. Baranov, R. A. Hovhannisyan, O. V. Skryabina, I. A. Golovchanskiy, V. I. Chichkov, C. Brun, T. Cren, V. M. Krasnov, A. A. Golubov, D. Roditchev, and V. S. Stolyarov, Local Josephson vortex generation and manipulation with a magnetic force microscope, \textit{ Nat. Commun.} {\bf 10}, 4009 (2019).

\bibitem{Correa_2019} A. Correa, F. Mompe\'{a}n, I. Guillam\'{o}, E. Herrera, M. Garcia-Hern\'{a}ndez, T. Yamamoto, T. Kashiwagi, K. Kadowaki, A. I. Buzdin, H. Suderow, and C. Munuera, Attractive interaction between superconducting vortices in tilted magnetic fields, \textit{ Commun. Phys.} {\bf 2}, 31 (2019)

\bibitem{Zhuravel_2006} A. P. Zhuravel, A. G. Sivakov, O. G. Turutanov, A. N. Omelyanchouk, S. M. Anlage, A. Lukashenko, A. V. Ustinov, and D. Abraimov, Laser scanning microscopy of HTS films and devices (Review Article), \textit{ Low Temp. Phys.} {\bf 32}, 592 (2006).

\bibitem{Koelle_2017} M. Lange, S. Gu\'{e}non, F. Lever, R. Kleiner, and D. Koelle, A high-resolution combined scanning laser and widefield polarizing microscope for imaging at temperatures from 4 K to 300 K, \textit{ Rev. Sc. Instr.} {\bf 88}, 123705 (2017).

\bibitem{Galin_2020} M. A. Galin, F. Rudau, E. A. Borodianskyi, V. V. Kurin, D. Koelle, R. Kleiner, V. M. Krasnov, and A. M. Klushin, Direct Visualization of Phase-Locking of Large Josephson Junction Arrays by Surface Electromagnetic Waves, \textit{ Phys. Rev. Applied} {\bf 14}, 024051 (2020).

\bibitem{Clem_1980} J. R. Clem and R. P. Huebener, Application of low-temperature scanning electron microscopy to superconductors, \textit{ J. Appl. Phys.} {\bf 51}, 2764 (1980).

\bibitem{Doenitz_2007} D. Doenitz, R. Kleiner, D. Koelle, T. Scherer, and K. F. Schuster, Imaging of thermal domains in ultrathin NbN films for hot electron bolometers, \textit{ Appl. Phys. Lett.} {\bf 90}, 252512 (2007).

\bibitem{Rosticher_2010} M. Rosticher, F. R. Ladan, J. P. Maneval, S. N. Dorenbos, T. Zijlstra, T. M. Klapwijk, V. Zwiller, A. Lupa\c{s}cu, and G. Nogues, A high efficiency superconducting nanowire single electron detector, \textit{ Appl. Phys. Lett.} {\bf 97}, 183106 (2010).

\bibitem{Giovati_2012} G. Ciovati, S. M. Anlage, C. Baldwin, G. Cheng, R. Flood, K. Jordan, P. Kneisel, M. Morrone, G. Nemes, L. Turlington, H. Wang, K. Wilson, and S. Zhang, Low temperature laser scanning microscopy of a superconducting radio-frequency cavity, \textit{ Rev. Sc. Instr.} {\bf 83}, 034704 (2012).

\bibitem{fietz1969hysteresis}  W.A. Fietz, and W.W. Webb. Hysteresis in superconducting alloys-Temperature and field dependence of dislocation pinning in niobium alloys. {\em Physical Review},{\bf 178}, 657 (1969).

\bibitem{park1992vortex} G.S. Park, C.E. Cunningham, B. Cabrera,  and  M.E. Huber. Vortex pinning force in a superconducting niobium strip. {\em Physical review letters}, {\bf 68}, 1920 (1992).

\bibitem{dasgupta1978flux} A. DasGupta, C.C. Koch, D.M. Kroeger, and Y.T. Chou. Flux pinning by grain boundaries in niobium bicrystals. {\em Philosophical Magazine B},{\bf 38}, 367-380 (1978).

\bibitem{Hovhannisyan_2021} Hovhannisyan, R. A., Grebenchuk, S. Yu., Baranov, D. S., Roditchev, D., \& Stolyarov, V. S., Lateral Josephson Junctions as Sensors for Magnetic Microscopy at Nanoscale. {\em J. Phys. Chem. Lett.} {\bf 12}, 12196 (2021).

\bibitem{stolyarov2022revealing} V. S. Stolyarov, V. Ruzhitskiy, R. A. Hovhannisyan, S. Grebenchuk, A. G. Shishkin, O. V. Skryabina, I. A. Golovchanskiy, A. A. Golubov, N. V. Klenov, I. I. Soloviev, et al., Revealing Josephson vortex dynamics in proximity junctions below critical current. {\em Nano letters} {\bf22}, 5715 (2022).

\bibitem{pinto2018dimensional} N. Pinto, S.J. Rezvani, A. Perali,  L. Flammia, M.V. Milosevic, M. Fretto, C. Cassiago,  and N. De Leo, . Dimensional crossover and incipient quantum size effects in superconducting niobium nanofilms. {\em Scientific reports}, {\bf8}, 4710 (2018).

\bibitem{Gubin} A.I. Gubin, K.S. ll$'$in,  S.A. Vitusevich,  M. Siegel, and  N. Klein,. Dependence of magnetic penetration depth on the thickness of superconducting Nb thin films.{\em Physical Review B}, {\bf 72}, 064503 (2005).










\bibitem{Mironov_2004} A.A. Fraerman, B.A. Gribkov, S.A. Gusev, V.L. Mironov, N.I. Polushkin and S.N. Vdovichev. Observation of MFM tip-induced remagnetization effects in elliptical ferromagnetic nanoparticles.{\em PHYSICS OF LOW DIMENSIONAL STRUCTURES}, {\bf 1/2}, 117-122 (2004).

\bibitem{di2019quantitative}  Di Giorgio, C., Scarfato, A., Longobardi, M., Bobba, F., Iavarone, M., Novosad, V., Karapetrov, G. \& Cucolo, A. M. Quantitative magnetic force microscopy using calibration on superconducting flux quanta. {\em Nanotechnology} {\bf 30}, 314004 (2019).

\bibitem{Zeinali_2016} A. Zeinali, T. Golod, and V. M. Krasnov, Surface superconductivity as the primary cause of broadening of superconducting transition in Nb films at high magnetic fields, \textit{ Phys. Rev. B} {\bf 94}, 214506 (2016).

\bibitem{Pearl_1964} J. Pearl, Current distribution in superconducting films carrying quantized fluxoids. {\em Appl. Phys. Lett.} {\bf 5}, 65 (1964).

\bibitem{Pearl_1966} J. Pearl, Structure of Superconductive Vortices near a Metal-Air Interface {\em J. Appl. Phys.}, {\bf 37}, 4139 (1966).

\bibitem{Ivanchenko-pla90} Yu.M.Ivanchenko and T.K.Soboleva, Nonlocal interaction in Josephson junctions {\em Phys.Lett.A} {\bf 147},65 (1990).

\bibitem{Mints-prb94} R. G. Mints and I. B. Snapiro,  Dynamics of Josephson pancakes in layered superconductors, {\em Phys. Rev. B} {\bf 49}, 6188 (1994).

\bibitem{Mints-prb95} R. G. Mints and I. B. Snapiro, Electromagnetic waves in a Josephson junction in a thin film, {\em Phys. Rev. B}
{\bf 51}, 3054 (1995).

\bibitem{Kogan-prb01} V. G. Kogan, V. V. Dobrovitski, J. R. Clem, Y. Mawatari, R. G. Mints, Josephson junction in a thin film {\em  Phys. Rev. B} {\bf 63}, 144501 (2001).

\bibitem{Krasnov-prl13} A. A. Boris, A. Rydh, T. Golod, H. Motzkau, A. M. Klushin, and V. M. Krasnov, Evidence for Nonlocal Electrodynamics in Planar Josephson Junctions {\em Phys. Rev. Lett}. {\bf 111}, 117002 (2013).

\bibitem{Samokhvalov-jetpl03} A. V. Samokhvalov, Maximal supercurrent of a Josephson junction in a field of magnetic particles {\em  JETP Letters} {\bf 78}, 3691 (2003).

\bibitem{Krasnov-prl10} T. Golod, A. Rydh, and V. M. Krasnov, Detection of the Phase Shift from a Single Abrikosov Vortex, {\em Phys. Rev. Lett.} {\bf 104}, 227003 (2010).

\bibitem{Clem-prb11} John R. Clem, Effect of nearby Pearl vortices upon the Ic versus B characteristics of planar Josephson junctions  in thin and narrow superconducting strips {\em Phys. Rev. B} {\bf 84}, 134502 (2011).

\bibitem{Kogan-Mints-prb14} V. G. Kogan, R. G. Mints, Interaction of Josephson junction and distant vortex in narrow thin-film superconducting strips {\em Phys. Rev. B} {\bf 89}, 014516 (2014).

\bibitem{Mironov-prb17} S. Mironov, E. Goldobin, D. Koelle, R. Kleiner, Ph. Tamarat, B. Lounis, and A. Buzdin, Anomalous Josephson effect controlled by an Abrikosov vortex {\em Phys. Rev. B} {\bf 96}, 214515 (2017).

\bibitem{Krasnov-prb19} T. Golod, A. Pagliero, and V. M. Krasnov,  Two mechanisms of Josephson phase shift generation by an Abrikosov vortex, {\em Phys. Rev. B} {\bf 100}, 174511 (2019).

\bibitem{Abramowitz-Stegun} Handbook of Mathematical Functions, edited by M. Abramowitz and A. Stegun ~U.S. GPO, Washington, D.C., (1965).




\end{thebibliography}

\section{Acknowledgements}

The MFM experiments were carried out with the support of the Russian Science Foundation project No. 23-72-30004 (https://rscf.ru/project/23-72-30004/). The samples were elaborated owing to the supported by the Ministry of Science and Higher Education of the Russian Federation (No.FSMG-2023-0014).

\section{Author contributions}

V.S.S. suggested the idea of the experiment; V.S.S. conceived the project and supervised the experiments; R.A.H., S.Yu.G., A.G.Sh., O.V.S., N.M.L., I.A.G. ... and V.S.S, performed the sample and surface preparation for MFM experiments; D.R., I.A.G., and V.S.S. provided the explanation of the observed effects; A.S.M. and A.S. did numerical modelling; D.R. and V.S.S. wrote the manuscript with the essential contributions from other authors.



\end{document}